\documentclass[12pt]{article}
\newcommand{\beq}{\begin{equation}}
\newcommand{\eeq}{\end{equation}}
\newcommand{\bea}{\begin{eqnarray}}
\newcommand{\eea}{\end{eqnarray}}

\newcommand{\bra}{\langle}
\newcommand{\ket}{\rangle}

\usepackage{graphicx}
 \usepackage{setspace}
 \usepackage{amsfonts}
 \usepackage{amsmath}  
 \usepackage{txfonts} 

\begin{document}
\pagestyle{empty}
\begin{center}
\vskip 2cm
{\bf \LARGE Does Entropic Gravity Bound the Masses of the Photon and Graviton?}
\vskip 1cm
J. R. Mureika\footnote{Email: jmureika@lmu.edu} \\

{\small \it Department of Physics, Loyola Marymount University, Los Angeles, CA}

and\\

R. B. Mann\footnote{Email: rbmann@sciborg.uwaterloo.ca}

{\small \it Perimeter Institute for Theoretical Physics\\ and\\ Department of Physics and Astronomy, University of Waterloo, Waterloo, ON~~Canada}
\end{center}

{\noindent{\bf Abstract} \\
If the information transfer between test particle and holographic screen in entropic gravity respects both the uncertainty principle and causality, a lower limit on  the number of bits in  the universe relative to its mass may be derived.  Furthermore, these limits indicate particles that putatively travel at the speed of light -- the photon and/or graviton --  have a non-zero mass $m \geq 10^{-68}~$kg. This result is found to be in excellent agreement with current experimental mass bounds on the graviton and photon, suggesting that entropic gravity may be the result of a (recent) softly-broken local symmetry. 
Stronger bounds emerge from consideration of ultradense matter such as neutron stars, yielding limits of $m \geq 10^{-48}-10^{-50}~$kg, barely  within the experimental photon range and outside that of the graviton.  We find that for black holes these criteria cannot be satisfied, and suggest some possible implications of this result.
}


\section{Introduction}
The recent proposal \cite{padman1,padman2,verlinde} that gravity is an emergent phenomenon of entropy maximization has added to a growing list of gravitational-thermodynamic dualities \cite{beckenstein1,beckenstein2,beckenstein3,hawking,jacobson,thooft,susskind,maldacena,bousso}, raising much interest
amongst the theoretical physics community.   It can be understood as an application of the holographic principle, which states that there is a duality between a physical description (including
gravity) of a volume of space  and  a corresponding (non-gravitational) physical theory formulated on the boundary this volume.  

While the motivation behind studying these
dualites stems from a desire to better understand quantum gravity, cosmological considerations of holographic duality have been of some interest for a number of years \cite{dsCFT1,dsCFT2,dsCFT3}.  Further progress came with with derivation of the
Friedmann equations from the first law of thermodynamics \cite{frw} on the apparent cosmic event
horizon, along with the assumptions that the entropy is proportional to its area and temperature
to its surface gravity.   The advent of entropic gravity \cite{verlinde} has prompted theorists to deploy this novel framework not only in quantum gravity \cite{smolin,modesto,zhao} and quantum information \cite{myung,lee2,lee3},  but also to explain  a wide range of  cosmological phenomena, including implications for black hole temperature \cite{bh1,bh2,bh3,ohta2}, dark energy \cite{smoot1,lee,danielsson,Li:2010cj,ohta1}, and  inflation \cite{smoot2,cai,Li:2010bc,Wang:2010jm}. 

Entropic gravity, and hence holography, thus plays a deep role in connecting these two bookend-realms of universal scales.  Cosmological holography is essentially the proposal that all of the information in our universe is encoded in a structure on its cosmological horizon.  An application of this proposal has 
been put forth in a recent conjecture by Smoot \cite{smoot3}, that all possible past and future histories of the universe  are encoded on its apparent horizon, thereby making a connection.  

Here we proceed  along similar lines, asking how much the ``whole'' -- the total mass and information content of the Universe -- can tell us about the ``parts'' -- the lightest possible mass of elementary particles.  Specifically we point out that there is a lower limit to the number of bits of a holographic screen in entropic gravity provided the information transfer between a test particle and the screen respects both the uncertainty principle and causality.  When applied to the entire universe  ({\it i.e.} taking the holographic screen to be the boundary of the visible Universe), this   limit indicates that all quanta have a non-zero mass  $m \geq 10^{-68}~$kg. This implies that the exact symmetries governing the behaviour of photons and gravitons are broken, albeit very softly.  This lower bound is only a few orders of magnitude below current experimental bounds on the masses of these particles.   It is also weakly time-dependent on cosmic time scales, suggesting  possible tests  of cosmological holography and entropic gravity.

We also consider implications of this bound for holographic screens due to ultradense matter (such as neutron stars) and black holes.   For both neutron stars and stellar-mass black holes  the bound is larger but remarkably consistent at $m \geq 10^{-48}-10^{-50}~$kg.  While barely within the experimental range of the photon mass, such a value excludes current inferred bounds on the mass of the graviton.   This result seems to suggest that either the method discussed herein is either not applicable to the graviton (or perhaps to gauge particles in general), or alternatively that the graviton is not an actual particle in the entropic gravity scenario.  Taking a black hole to be the source of the holographic screen, we find that the bound is inversely proportional to the black hole radius. Hence for a sufficiently small black hole (smaller than about a solar mass) the bound cannot be empirically satisfied.   If massless quanta cannot be accommodated within entropic gravity, then this suggests a minimal bound on the mass of a black hole.

\section{Entropic Gravity Primer}
\label{entgravprimer}

An entropic force is one that drives a system's entropy to increase \cite{verlinde},
\beq
F_{\rm entropic} = T\frac{\Delta S}{\Delta x}~~.
\label{entforce}
\eeq
The foundation of the idea \cite{verlinde}  that gravitation is such a force relies on a holographic argument relating the entropy to the area of a screen, the temperature to the acceleration of the particle, and the thermodynamic equipartition theorem.  A test particle of mass $m$ is located some distance from another (presumably larger) mass $M$, the latter of which generates a holographic screen at a distance $r$.  On this screen, the holographic information from mass $M$ is encoded as 
\beq
N = \frac{A}{\ell_P^2} = \frac{4\pi r^2c^3}{G\hbar}
\label{bits}
\eeq
 bits.  As the mass $m$ approaches the screen, its own entropy bits begin to transfer to the screen, and it is this increase in screen entropy that generates an attractive force.

Specifically, the entropy transferred by $m$ at a distance $\Delta x$ is
\beq
\Delta S = 2\pi k_B \frac{\Delta x}{\lambdabar}~~,~~\lambdabar = \frac{h}{mc}
\label{deltas}
\eeq
so that a ``quantized'' unit of entropy $\Delta S = 2\pi k_B$ is incremented when the particle is within a distance equal to its Compton wavelength.  The energy on the screen obeys thermal equipartition,
\beq
E = Mc^2 =  \frac{1}{2}N k_B T~~~\longrightarrow~~~k_B T = \frac{2Mc^2}{N}
\label{equip}
\eeq
Substituting (\ref{bits}) and (\ref{deltas}) into (\ref{entforce}) yields Newton's universal law of gravitation,
\beq
F_{\rm entropic} = - \frac{GmM}{r^2}
\eeq
which carries with it the novelty that the force is emergent instead of fundamental.\footnote{The minus sign signifies an attractive force,  which results from the sign of the term $\Delta x$ used in the calculation of the entropic force actually being negative \cite{verlinde}.}

\section{Bounding Information Transfer}
\label{speed}

The formulation of the entropy mechanism suggests that there is an inherent uncertainty $\Delta x$ in the location of the test mass $m$ relative to the holographic screen.  Indeed, when $\Delta x \sim \lambda_c$, the entropy of $m$ merges with that of the screen.  This leads one to suggest that, when the position uncertainty of $m$ is $\Delta x$, there is a statistical fluctuation of the screen's entropy $\Delta S$, and hence an uncertainty in its energy $\Delta E \sim T \Delta S$ that must abide by quantum mechanical considerations\footnote{Several authors have previously addressed the connection between entropic gravity and generalized position-momentum uncertainty principles \cite{vancea,ghosh}.}

A naive limit on the variation in momentum of the mass $m$ may be derived from the standard Heisenberg Uncertainty Principle (HUP),
\beq
\Delta p \geq \frac{\hbar}{2\Delta x} = \frac{mc}{4\pi}~~.
\label{hup}
\eeq
The uncertainty in the velocity is thus $\Delta v \geq \frac{c}{4\pi}$, but is really inconsequential to the problem at hand as it is the {\it average} velocity that must respect $\bra v \ket <c$.   Furthermore, the uncertainty principle further suggests that the fluctuation in the energy of the screen is constrained to occur during the interval
\beq
\Delta E ~\Delta t \geq \frac{\hbar}{2} ~~\Longrightarrow~~~\Delta t \geq \frac{\hbar}{\Delta 2E}
\label{teu}
\eeq
Once the particle is within a Compton wavelength of the screen, the ``speed'' of information transfer is on the order of $v_I \sim \frac{\Delta x}{\Delta t}$, which according to Equation~\ref{teu} becomes
\beq
v_I \sim \frac{\Delta x}{\Delta t} \leq \frac{2\lambdabar \Delta E}{ \hbar}
\eeq
Following the rationale applied in Section~\ref{entgravprimer}, the uncertainty in the energy may be written
\beq
\Delta E = T\Delta S = 2\pi k_B T = \frac{4\pi Mc^2}{N}
\eeq
and thus
\beq
v_I \leq \left(\frac{16\pi^2}{N}\cdot \frac{M}{m}\right) c
\label{vinf1}
\eeq
Imposing a causality bound on this upper limit yields
\beq
\left(\frac{16\pi^2}{N}\cdot \frac{M}{m}\right) c~ \leq~ c
~~~\Longrightarrow~~~
\left(\frac{16\pi^2}{N}\cdot \frac{M}{m}\right)~ < ~1
 \label{caus1}
\eeq

This naive approach is fraught with ambiguity, however, due to the well-known fact that time itself is a parameter and not associated with any hermitian operator.  A clarification was offered by Mandelstam and Tamm \cite{mantamm}, who modified eq. (\ref{teu})  via an auxiliary observable $O$, obtaining
\beq
\Delta E \left(\frac{\Delta O}{\left|\frac{d\bra O\ket}{dt}\right|}\right) \geq \frac{\hbar}{2}~~.
\label{mtu}
\eeq
relating the standard deviation of the energy operator of some non-stationary state to
 the standard deviation ${\Delta O}$.  Identifying $O \rightarrow X$ as the position operator
 yields the group velocity 
$\frac{d\bra X \ket}{dt} = \bra v\ket $ of the particle's wavepacket,
which can be associated with the speed of information transfer from the wavepacket to the screen.
Equation~(\ref{mtu}) may thus be recast as a constraint on $\bra v_I\ket$,
\beq
\bra v_I \ket \leq \frac{2\Delta E~\Delta X}{\hbar} = \frac{2\lambdabar T\Delta S}{\hbar} = \frac{4\pi k_B T\lambdabar}{\hbar}
\eeq

Inserting equations ~(\ref{deltas},\ref{equip}) in the above relation and noting $E = Mc^2$ for the mass associated with the screen gives the bound
\beq
\bra v_I \ket \leq \frac{16\pi^2}{N} \frac{M}{m} c 
\label{vinf2}
\eeq
We impose a strict causality relation by demanding that this upper bound is always less than
the speed of light, in which case
\beq
 \frac{16\pi^2}{N} \frac{M}{m} \leq 1
 \label{caus2}
\eeq
which is an identical result to that obtained previously (\ref{vinf1}).

\section{Bounds on $m$ in the limit $\bra v \ket = c$}

The above result (\ref{caus2}) has deep implications for the underlying framework of entropic gravity.    It implies a relationship between the number of bits $N$ on the holographic screen generated by a mass $M$ and any test mass $m$  in its vicinity.  
Re-expressed as the bound $Nm \geq 16\pi^2 M$,  one concludes the information on the screen is affected not just by its ``generator'' (or source) $M$, but also by $m$.  One can interpret this as indicating that any holographic screen has a minimal number of bits, dependent on the ratio $M/m$.    We propose that this relationship implies a lower bound on the mass of any quanta.

\subsection{Bounds from Cosmology}

Although the Standard Model and General Relativity predict massless photons and gravitons, respectively, the possibility exists that both particles may indeed have some finite, non-zero rest mass.  Proca demonstrated that the addition of a photon mass could be realized in a Lorentz-invariant manner \cite{proca}, via the Lagrangian
\beq
{\cal L} = -\frac{1}{4}F_{\mu \nu} F^{\mu \nu}-\frac{1}{2}\frac{m^2c^2}{\hbar^2} A_\mu A^\mu~~.
\label{procalag}
\eeq
The addition of mass to the photon would introduce a frequency-dependent dispersion relationship, and also modify Maxwell's equations in an ultimately testable fashion.  Experimental considerations have constrained $m_\gamma < 10^{-49}-10^{-54}~$kg \cite{massbounds}.  A massive graviton would possess a similar dispersion relation \cite{will}, which would manifest itself as signal arrival-time delays (or even inversions) in gravity wave detectors \cite{massivegraviton}.   Such a property could also be used to model long-range deviations from general relativity, and hence provide an explanation for galaxy rotation curves and late-time cosmological acceleration.  LIGO/VIRGO and LISA-scale gravitational wave searches from {\it e.g.} massive compact binary coalescence could potentially constrain the graviton's Compton wavelength to be $\lambdabar_g \geq 10^{12}-10^{16}~$km, respectively, yielding an upper mass limit of about $m_g \leq 10^{-58}-10^{-62}~$kg \cite{will}.  Other theoretical and model-dependent considerations provide similar estimates in the range $m_g \sim 10^{-55}-10^{-69}~$kg (see Table~\ref{masstable} for a summary).

It is therefore of interest to know, in the spirit of Smoot's thesis \cite{smoot3}, what one can learn from the entropic connection between the ``whole'' -- total information content of the universe ({\it i.e.} total mass)-- and the ``parts'' -- the particles.  The inequality  (\ref{caus2}) may alternatively be written as one bounding the mass $m$ in terms of the holographic screen and $M$.  If we take $M$ be the
mass of the universe, $M_U$ and $N=N_U$ to be the number of events or operations that could have occurred within the age $T_U$ and size of the universe $R_U$ we obtain
\beq
m \geq \frac{16\pi^2 M_U}{N_U}
\label{massbound}
\eeq

This yields the startling implication that all quanta --
including ``light-like'' particles such as photons and gravitons -- actually possess a negligible but manifestly non-zero mass. Can this relationship thus be used to pin down the mass bounds on such particles?

Applying the holographic framework to the contents of the entire universe, one can write $N \sim 10^{122}$ for a sphere of radius $r \sim H^{-1}$ \cite{SLloyd}, a number comparable to the ratio of the area of the apparent horizon of the universe to the Planck area \cite{smoot3}.
An estimate of the visible mass content of the universe is on the order of $M_{\rm vis.} \sim 10^{52}~$kg, and including additional contributions from dark sources increases this by just under two orders of magnitude.   We may also approximate the mass of the universe from the critical density, whose value is roughly $\rho_c = 3c^2H^2/8\pi G \simeq 10^{-30} -10^{-29}~$g$/$cm$^3$, depending on the value of $H$.   
Taking the age of the Universe to be 13.7~Gyr, one can  gauge  its ``size" ({\it i.e.} that of a co-moving sphere) to be $R\simeq 4\times 10^{26}$~m.  This implies a mass of roughly $M \sim 10^{54}~$kg.  Current data from WMAP indicates a baryon and dark matter density of 4.56\% and 22.7\%, respectively \cite{wmap}.  It is unclear whether or not dark energy should be included in this calculation, however, since it is unclear whether or not it will contribute to the holographic information.  This omission does not significantly alter our conclusions, however, since it will change the result by less than an order of magnitude. 

In this case, the holographic screen is the bounding surface of the visible Universe, {\i.e.} a co-moving sphere of radius $R=R_U$. Based on the range $M \sim 10^{52}-10^{54}~$kg, the inequality (\ref{massbound}) therefore gives a numerical value
\beq
m_{\rm min} \sim (10^{-68}-10^{-66})~{\rm kg}~~.
\label{mmin}
\eeq

This range represents the smallest non-zero mass for any particle quanta  in the entropic gravity framework.  Remarkably, the range of values is quite commensurate with the experimental bounds cited in Table~\ref{masstable}.   One is tempted to conclude that the non-zero rest mass for heretofore-thought massless particles has come about due to some kind of (recent) soft symmetry breaking in the appropriate sector. 
\begin{table}[h]
{\begin{center}
\begin{tabular}{cc|cc}\hline
\multicolumn{2}	{c|}{Photon} &\multicolumn{2}{c}{Graviton}\\ \hline \hline
 Source & $m_\gamma$ (kg)& Source & $m_g$ (kg) \\ \hline
 Coulomb's law & $2\times 10^{-50}$ & Gravitational wave dispersion & $10^{-55}$ \\
Jupiter's magnetic field & $7\times 10^{-52}$ & Pulsar timing & $2\times 10^{-59}$ \\
Solar wind magnetic fields & $2\times 10^{-54}$ & Gravity over cluster sizes & $2\times 10^{-65}$ \\
 Cosmic magnetic fields&  $10^{-62}$ & DGP constraints & $10^{-67}-10^{-69}$ \\ \hline
\end{tabular}
\caption{Theoretical and experimental photon and graviton limits from various sources (adapted from \cite{massbounds}).}
\label{masstable}
\end{center}}
\end{table}

Application of the entropic gravity formalism to massless particles is somewhat problematic, in part due to the difficulty of their localization relative to a holographic screen.  A recent suggestion
 \cite{xgh} for incorporating photons in entropic gravity involves positing that one can substitute 
 $E/c^2$ in place of $m$, where $E$ is the energy of the photon.  In this context
the effective mass $m_\gamma$ of the photon obeys a force law of the form
\beq
F_{\rm entropic} = -\frac{GM_{\rm BH}}{r^2} \frac{E_\gamma}{c^2}, ~~~E_\gamma = m_\gamma c^2
\eeq
We posit instead that a causality- and quantum mechanical-respecting entropic framework necessitates non-zero masses for all particles.
The effective mass derived in \cite{xgh} yields photon  masses of $m_\gamma \sim (10^{-30} - 10^{-45})~$kg for photons covering the energy spectrum of $E_\gamma \sim 10^{-15} - 1~$MeV.   This result is somewhat tenuous, however, as it implies a variable (energy-dependent) photon mass that is not commensurate with standard particle theory\footnote{Variable-mass quanta have been discussed in the literature, most recently as a consequence of conformal symmetry presevration in unparticle physics \cite{unparticle}}. 

It is also worth noting that our bound (\ref{massbound}) is comparable to, but distinct from, the mass of a particle whose Compton wavelength is the size of the observable universe $R_U$.   This latter quantity is given by 
$m_c =  \frac{h}{R_U c}= 1.7 \times10^{-68}$ kg, whereas the bound (\ref{massbound}) is
\beq
m \geq \frac{16\pi^2 M_U}{N_U} = \frac{16\pi^2 M_U \ell_p t_p}{R_U T_U} = \frac{16\pi^2 GM_U \hbar}{R_U T_U c^4}
\eeq
 which yields the comparable but distinct value (\ref{mmin}),  upon taking $T_U = 13.7~$Gyr  ($4.3\times 10^{17}~$sec)  and $R_U = 10^{26}~$m.

It is of interest to note that the expression (\ref{massbound}) is inherently time-dependent, since it depends on the size (age) of the universe.  One might ask, then, how the bound on the smallest mass was different in the past.  As an illustrative example, we consider the value at the surface of last scattering, where $z \approx 1100$ \cite{spergel}.  One may then approximate the ``size'' of the universe at recombination from the scale difference as $R_{\rm CMB} = 10^{-3} R_U$, and so the number of bits can thus be calculated as $N_{\rm CMB} = 10^{-6} N_U$.   If the mass of the universe has not significantly changed over this time, the lower limit (\ref{mmin}) increases by a factor of $10^6$, which is still within the acceptable experimental bounds.

At much earlier stages, the number of bits $N_U(r)$ grows significantly smaller, and one expects the limit (\ref{massbound}) to grow.  One would expect corrections to the area/entropy law to become more important during these eras, thus altering the holographic bounds derived in this paper.   It is an open question as to how to treat the mass content $M_U(r)$ at these earlier times, since the existence of a particle horizon may alter the estimate.  It is reasonable to assume that the ratio $M_U(r)/N_U(r)$ approaches a finite (possibly vanishing) value as $r\rightarrow 0$ due to quantum gravity corrections.

\subsection{Bounds from mass density}

Alternatively, the bound (\ref{massbound}) may be expressed in terms of the size of the source mass distribution and its density,
\beq
m \geq \rho R \ell_P^2
\label{densitybound}
\eeq
up to factors of order unity, using equation (\ref{bits}) and taking the screen to be infinitesimally close to the edge of the source mass.  Appealing to (\ref{densitybound}), one may consider the bounds imposed by the densest structures known in the Universe.  Neutron stars have densities on the order $\rho_{\rm NS} \sim 10^{17}~{\rm kg}\; {\rm m}^{-3}$ and radii $R_{\rm NS} \sim 10^3-10^4~$m.  Applying these values to the inequality (\ref{densitybound}) gives the limit
\beq
m \geq 10^{-50}~{\rm kg}~~.
\eeq
This is in reasonable agreement with the upper limit on photon mass listed in Table~\ref{masstable}, and  is stable over the age of the Universe (barring time-dependence of the fundamental constants).   

Of course the above bounds are incompatible with the semi-empirical bounds on the graviton mass given in Table~1.  This can be interpreted in a number of ways.  One possibility is that the graviton is not an actual particle in the entropic gravity scenario, indicating that excitations of the gravitational field need a qualitatively different description within this context.  Another possibility is that our causality bounds are evaded for some reason by gravitons.   A third possibility is that all massless quanta need a different descriptive explanandum within entropic gravity.  This last point will become more pertinent when we consider black holes as the source masses.

\subsection{Bounds from black hole horizons} 

An alternative approach is to consider the causality-preserving mass bound imposed on a test particle approaching the horizon of a black hole, where the area-entropy relationship is saturated.  Since $M_{\rm BH} = \frac{R_H c^2}{2G}$ and $N = \frac{4\pi R_H^2}{\ell_P^2}$, we find that the bound is no longer mediated by a balance between density and scale size, but rather scales purely as the inverse of the horizon size of the source,
\beq
m \geq \frac{\hbar}{8\pi c R_H}
\label{bhbound}
\eeq
 again in the limit that the screen is infinitesimally close to the horizon.
We see that the bound grows as $R_H^{-1}$, and thus arbitrarily small black holes will necessitate arbitrarily large values for $m_{\rm min}$. There are several ways of interpreting this result.

For a stellar-mass black hole having $R_H \sim 10^4~$m, the corresponding bound is
\beq
m \geq 10^{-48}~{\rm kg}~~,
\eeq
which is two orders of magnitude larger than the secure bound  \cite{massbounds} of $10^{-50}~{\rm kg}$ for the photon.   Hypothetical primordial black holes of initial mass $M_{\rm PBH} \sim 10^{12}~$kg would yield $m \geq 10^{-28}~$kg, but assuming the standard evaporation process may today be as small as $10^{-9}~$kg \cite{pbh}.  These bounds are clearly unacceptable for the graviton and photon.  

An alternative, then, would be to abandon the applicability of equation (\ref{caus2}) for gauge fields, assuming they can be accounted for in the context of entropic gravity by other means. Re-expressing the relation (\ref{bhbound})   as a constraint on the magnitude of the horizon radius in terms of the test mass' Compton wavelength 
\beq
R_H \geq \frac{\lambdabar}{8\pi}
\eeq
implies a lower bound on the mass of any black hole, on the order $\lambdabar^{-1}$ for the lightest quanta.  The extreme bound may be obtained from the lightest neutrino, whose mass\footnote{Although individual neutrino masses are not measured, the measured value of the mass-squared difference between flavors is in the range $\Delta m^2 \in (10^{-7}-10^{-3})~eV^2$ \cite{pdg}.  We assume the natural hierarchy $m_{\nu_1} \gg m_{\nu_i}~(i =2,3)$.} is approximately $m_\nu \sim 10^{-3}-1~$eV$=10^{-35}-10^{-32}$~kg. The minimal bound on horizon radius is thus
\beq
R_H \geq 10^{-7}~{\rm metres}
\eeq
which is clearly satisfied by current observations.

One can either conclude that the entropic gravity formalism breaks down in the (quantum) black hole region, or that to respect causality it is incompatible with our current understanding of gravitation.

\section{Conclusions}

In summary, we have shown that entropic gravity suggestively bounds the  minimum  mass of any quanta, if causality and the uncertainty principle are upheld.   The derived bounds are consistent with the photon, but depending on the choice of holographic source, the method does not account for the observed graviton mass bounds.  We suggest that this is either a failure of the entropic gravity formalism, or alternatively that the mass bound is not applicable to gauge fields in general.  

We theorize this effect, and by proxy the basis of entropic gravity, has arisen due to a spontaneously broken symmetry.  While the causality bounds have been imposed for bits traveling over distances comparable to the Compton wavelength
 of the test mass $m$, it is unclear how this will influence the flow of information over macroscopic distances.  Presumably when the separation $\Delta x$ is sufficiently large, the information transfer speed is again limited, and thus distant screens cannot ``know'' about approaching particles.  This might suggests that gravity is actually a local phenomenon, {\it i.e.} interactions between a particle and a nearby screen.  Each screen thus acts as a gravitational ``relay'' that transmits information to the next.  

Overall, our findings further the suggestions of previous authors \cite{smoot1,smoot2,smoot3} that the largest and smallest facets of our universe are implicitly connected via this new ``duality,'' opening exciting prospects for applications of emergent gravity to the quantum regime.

\vskip 2cm
\noindent{\bf Acknowledgments}\\
The authors would like to thank Niayesh Ashfordi and Leonardo Modesto for insightful discussions.  This work was supported by the Natural Sciences and Engineering Research Council of Canada (RBM) and the Research Corporation for Science Advancement (JRM).  Additionally, JRM acknowledges the generous hospitality of the University of Waterloo Department of Physics and Astronomy and the Perimeter Institute for Theoretical Physics, at which this research was partially conducted.

\end{document}